\documentstyle[stwol,epsfig]{article}
\def\({ \left( }
\def\){ \right) }
\def\b{\begin{equation}}
\def\e{\end{equation}}
\def\={\ =\ }

\def\+{\ +\ }
\def\-{\ -\ }

\def\Ls{\cal L \rm}

\def\mumu{$\mu^+\mu^-$}
\def\ee{$e^+e^-$}
\def\pp{$pp$}
\def\ppbar{$p\bar{p}$}

\begin{document}
\title{FUTURE COLLIDERS}
\author{R. B. PALMER and  J. C. GALLARDO }
\address{Brookhaven National Laboratory,  Upton, NY 11973-5000, USA}
\twocolumn[
\maketitle
\abstracts{The high energy physics advantages, disadvantages and luminosity 
requirements of hadron (\pp, \ppbar), of lepton (\ee, \mumu) and 
 photon-photon colliders are 
considered. Technical arguments for increased energy in each type of 
machine are presented. Their relative size, and the implications of size on 
cost are discussed. }]

\section{Physics Considerations}

\subsection{General}
Hadron-hadron colliders (\pp \, or \ppbar) generate interactions between  the
many constituents of the  hadrons (gluons, quarks and antiquarks); the initial 
states are not defined and most interactions occur at relatively  low energy
generating a very large background of uninteresting  events. The rate of the
highest energy events is higher for  antiproton-proton machines, but this is a
small effect for  colliders above a few TeV. In either case the effective
individual interaction energies are a relatively small fraction of the total
center of mass energy. Nevertheless, because high energy  hadron machines have
been relatively easier and cheaper to build,  and because all final states are
accessible, many initial  discoveries in Elementary Particle Physics have been
made with these machines. 

In contrast, lepton-antilepton and photon-photon colliders  generate
interactions between the fundamental point-like  constituents in their beams,
the reactions generated are  relatively simple to understand and there is no
background of low  energy events. If the center of mass energy is set equal to
the  mass of a suitable state of interest, then there can be a large  cross
section in the {\bf s}-channel, in which a single state is  generated by the
interaction. In this case, the mass and quantum  numbers of the state are
constrained by the initial beams. If the  energy spread of the beams is
sufficiently narrow, then precision  determination of masses and widths are
possible. 

A gamma-gamma collider also has well defined initial states, 
complementing those attainable with lepton colliders. 

For most purposes (technical considerations aside) \ee and \mumu  colliders
would be equivalent. But in  the particular case of {\bf s}-channel Higgs boson
production,  the cross section, being proportional to the mass squared, is 
more than 40,000 times greater for muons than electrons.  When technical
considerations are included, the situation is  more complicated. Muon beams are
harder to polarize and muon  colliders will have much higher backgrounds from
decay  products of the muons. On the other hand muon collider  interactions
will require less radiative correction and will have  less energy spread from
beamstrahlung.

Each type of collider has its own advantages and disadvantages 
for High Energy Physics: they are complementary.

\subsection{Required Luminosity for Lepton Colliders} 

   In lepton machines the full center of mass of the leptons is 
available for the final state of interest and the {\it effective 
energy} is equal to the total center of mass energy. 
 \b
E_{\rm eff}\ =\ E_{\rm c\,of\,m}
 \e

\begin{figure}[t!] 
\centerline{\epsfig{file=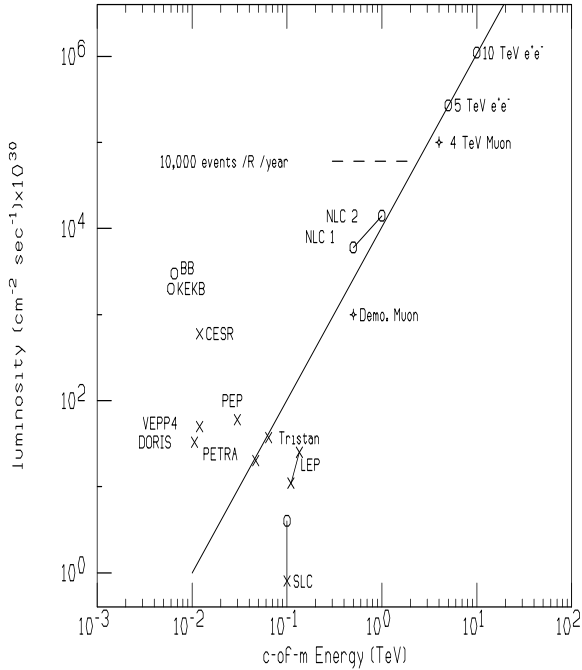,height=3.5in,width=3.0in}}
 \caption{Luminosity of lepton colliders as a function of energy} 
 \label{lumfig} \end{figure}

   Since fundamental cross sections fall as the square of the 
center of mass energies involved, so, for a given rate of events, 
the luminosity of a collider must rise as the square of its 
energy. A reasonable target luminosity is one that would give 
10,000 events per unit of R per year: 
 \b
\Ls_{req.}\ \approx \  10^{34}\ (cm^{-2} s^{-1})\  \( {E_{\rm eff} \over 1\ 
(TeV)} \)^2 \label{reqlum} 
 \e
   Fig.~\ref{lumfig} shows this required luminosity, together with crosses at 
the approximate achieved luminosities of some lepton colliders. Target 
luminosities of possible future colliders are also given as circles. 
\subsection{The Effective Energies of Hadron Colliders}
   Hadrons, being composite, have their energy divided between 
their several constituents. A typical collision of 
constituents will thus have significantly less energy than that 
of the initial hadrons. Studies done in Snowmass 82 and 96 
suggest that given the required luminosity (as defined in 
Eq.~\ref{reqlum}) the hadron machine's {\it effective energy} is 
about $1/10\,{\rm th}$ of its total: 
$$
E_{\rm eff}(\Ls=\Ls_{req.})\ \approx\ {E_{\rm c\,of\,m} \over 10}
$$
The same studies have also concluded that a factor of 10 in 
luminosity is worth about a factor of 2 in effective energy, this 
being approximately equivalent to: 
  $$
E_{\rm eff}(\Ls)\ \propto \ \left(\Ls /\Ls_{\rm req.}\right)^{0.3}
  $$
From which, with Eq.~\ref{reqlum}, one obtains:
 \b
E_{\rm eff}(TeV) \approx  
  \( {E_{\rm c\ of\ m}\over 10 (TeV)} \)^{0.6}
               \( {\Ls \over 10^{34}(cm^{-2}s^{-1})} \)^{0.2}
\label{Eeffeq}
 \e
 \section{Technical Considerations}
 \subsection{Hadron-Hadron Machines}


An antiproton-proton collider requires only one ring, compared  with the two
needed for a proton-proton machine, but the  luminosity of an antiproton-proton
collider is limited by the  constraints in antiprotons production. Luminosities
of $10^{33}\ {\rm cm}^{- 2}{\rm s}^{-1}$ may be achievable at FNAL with
antiproton-proton, but  LHC, a proton-proton machine, is planned to have a
luminosity  of $10^{34}\ {\rm cm}^{-2}{\rm s}^{-1}$, and might \cite{lumlim}
 be upgradable to $10^{35}\ {\rm cm}^{-2}{\rm s}^{-1}$. Radiation damage to a 
detector would, however, then be a severe problem. The 60 TeV Really Large 
Hadron Colliders (RLHC high and low fields ) discussed at Snowmass are  being
designed as proton-proton machines with luminosities of  $10^{34}\
{\rm cm}^{-2}{\rm s}^{-1}.$ 

The size of hadron-hadron machines is limited by the field of  the magnets used
in their arcs. A cost minimum is obtained when a  balance is achieved between
costs that are linear in length, and  those that rise with magnetic field. The
optimum field will  depend on the technologies used both for the the linear 
components (tunnel, access, distribution, survey, position  monitors,
mountings, magnet ends, etc) and those of the magnets  themselves, including
the type of superconductor used. 

The first hadron collider, the 60 GeV ISR at CERN, used  conventional iron pole
magnets at a field less than 2~T. The  only current hadron collider, the 2 TeV
TeVatron, at FNAL, uses NbTi  superconducting magnets at approximately
$4\,{}^\circ K.$ The  14 TeV Large Hadron Collider (LHC), under construction at
CERN,  plans to use the same material at $1.8\,{}^\circ K.$ 

\begin{figure}[t!] 
\centerline{\epsfig{file=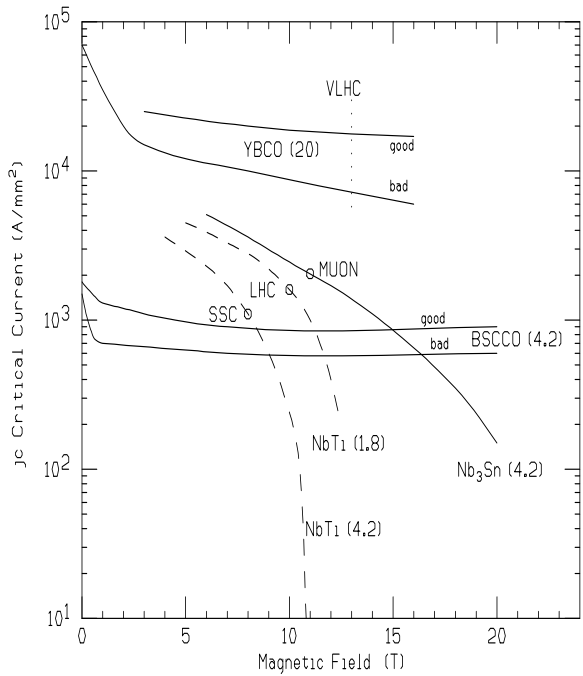,height=3.5in,width=3.0in}}
\caption{Critical current densities of superconductors as a function of
magnetic field.} \label{super} 
\end{figure}

Future colliders may use new materials allowing higher  magnetic fields.
Fig.\ref{super} shows the critical current densities of various superconductors
as a function of magnetic field. The numbers in parenthesis refer to the
temperatures in ${}^{\circ}$ K. {\it Good} and {\it bad} refer to the best and
worst performance according to the orientation in degree of the tape with
respect to the direction of the magnetic field. 
Model magnets have been made with Nb$_3$Sn, and  studies are underway on the
use of high T$_c$ superconductor.  {Bi$_2$Sr$_2$Ca$_1$Cu$_2$O$_8$} (BSCCO)
material is currently available in useful  lengths as powder-in-Ag tube
processed tape. It has a higher  critical temperature and field than
conventional superconductors,  but, even at $4\,{}^\circ K,$ its current
density is less than Nb$_3$Sn  at all fields below 15~T. It is thus unsuitable
for high  field accelerator magnets. In contrast YBa$_2$Cu$_3$O$_7$ (YBCO) 
material has a current density above that for Nb$_3$Sn ($4\,{}^\circ K$  ), at
all fields and temperatures below $20\,{}^\circ K.$  But this material must be
deposited on specially treated  metallic substrates and is not yet available in
lengths greater  than 1 m. It is reasonable to assume, however, that it will be 
available in useful lengths in the not too distant future.  
\begin{figure}[t!] 
\centerline{\epsfig{file=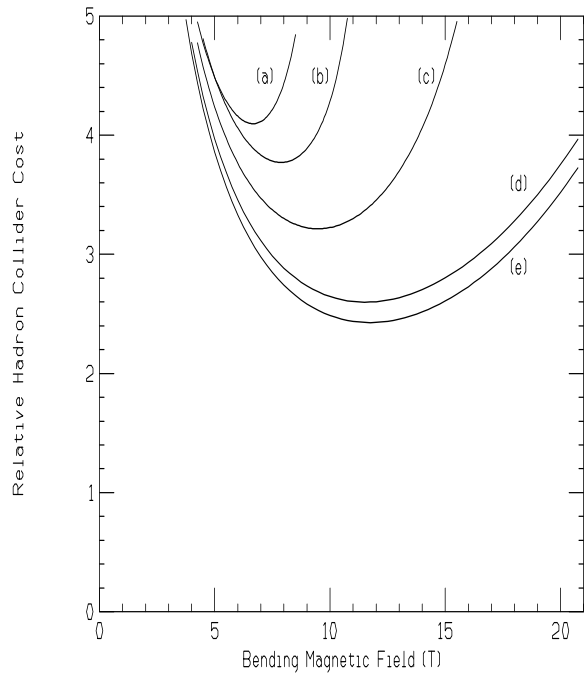,height=3.5in,width=3.0in}}
\caption{Relative costs of a collider as a function of its bending magnetic 
field, for different superconductors and operating temperatures} \label{mag} 
\end{figure} 

A parametric study \cite{magcost} was undertaken to learn what the use of  such
materials might do for the cost of colliders. 2-in-1 cosine  theta
superconducting magnet cross sections were calculated using  fixed criteria for
margin, packing fraction, quench protection,  support and field return.
Material costs were taken to be linear  in the weights of superconductor,
copper stabilizer, aluminum  collars, iron yoke and stainless steel support
tube. The  cryogenic costs were taken to be inversely proportional to the 
operating temperature, and linear in the outer surface area of  the cold mass.

The values of the cost dependencies were scaled from LHC  estimates. 
Results are shown in Fig.~\ref{mag}. Costs were calculated  assuming NbTi at
(a) $4\,{}^\circ K$, and (b) $1.8\,{}^\circ K$,  (c) Nb$_3$\,Sn at
$4.3\,{}^\circ K$, and (d) and (e) YBCO High T$_c$ at  $20\,{}^\circ K.$ 
NbTi  and Nb$_3$\,Sn  costs per unit weight were taken to be the  same; YBCO
was taken to be either equal to NbTi (in (d)), or 4 times NbTi (in (e)). 

It is seen that the optimum field moves from about 6~T for  NbTi at
$4\,{}^\circ K$ to about 12~T for YBCO at $20\,{}^\circ K$;  while the total
cost falls by almost a factor of 2; i.e., the optimized cost per  unit length
remains approximately constant. This might have been  expected: at the cost
minimum, the cost of linear and field  dependent terms are matched, and the
total remains about twice  that of the linear terms.  

It must be noted that the above study assumes a particular  type of magnet and
may not be indicative of the optimization for  radically different designs. A
group at FNAL \cite{pipe} is  considering an iron dominated, alternating
gradient, continuous,  single turn collider magnet design (Low field RLHC). Its
field would  be only 2~T and circumference very large (350 km for 60 TeV), but 
with its simplicity and with tunneling innovations it is hoped to  make its
cost lower than the smaller high field designs. There  are however greater
problems in achieving high luminosity with  such a machine than with the higher
field designs. 
 \subsection{Circular Electron-Positron Machines}

                 
Although the luminosities of most circular electron-positron  colliders has
been between $10^{31}$ and $10^{32}\,{\rm cm}^{-2}{\rm s}^{-1}$ (see Fig.\ref{lumfig}),  CESR is fast
approaching $10^{33}\,{\rm cm}^{-2}{\rm s}^{-1}$ and machines are  now being constructed with even high
values. Thus, at least in  principle, luminosity does not seem to be a
limitation, although  it may be noted that the 0.2~TeV electron-positron
collider LEP has a  luminosity below the above requirement.

At energies below 100 MeV, using a chosen reasonable bending  field, the size
and cost of a circular electron machine is  approximately proportional to its
energy. But at higher energies,  if the bending field $B$ is maintained, the energy lost
$\Delta  V_{\rm turn}$ to synchrotron radiation rises rapidly 
 \b
 \Delta  V_{\rm turn}\ \propto \ {E^4 \over R\ m^4}\ \propto \ {E^3\ B \over m^4}
\label{syncheq}
 \e
and soon becomes excessive ($R$ is the radius of the ring). A cost minimum is then obtained when the cost of
the ring is balanced by the cost of the  rf needed to replace the synchrotron
energy loss. If the ring  cost is proportional to its circumference, and the rf
is  proportional to its voltage then the size and cost of an  optimized machine
rises as the square of its energy. This  relationship is well demonstrated by
the parameters of actual  
machines (see Fig. \ref{length}). 

   The highest \ee collider is the LEP at CERN which has a 
circumference of 27 km, and will achieve a maximum center of mass 
energy of about 0.2 TeV. Using the predicted scaling, a 0.5 TeV 
circular collider would have to have a 170 km circumference, and 
would be very expensive. 

 \subsection{Electron-Positron Linear Colliders}
So, for energies much above that 0.2 TeV it is  impractical to build a circular
electron collider. The only  possibility is to build two electron linacs facing
one  another. Interactions occur at the center, and the electrons,  after they
have interacted, must be discarded. 

If the linacs are conventional, non-superconducting, structures, then there may
again be a cost trade off;  this time between the cost of rf to obtain
accelerating gradient,  and the linear costs of the structure, tunnel, etc. If
the  optimized gradient is less than its technical maximum, then the  cost per
unit length of an optimized machine should again be  about twice the linear
costs. But this time the {\it linear costs}  include the linac itself, and these
will be dependent on  technology and rf frequency. 

 \begin{figure}[t!] 
\centerline{\epsfig{file=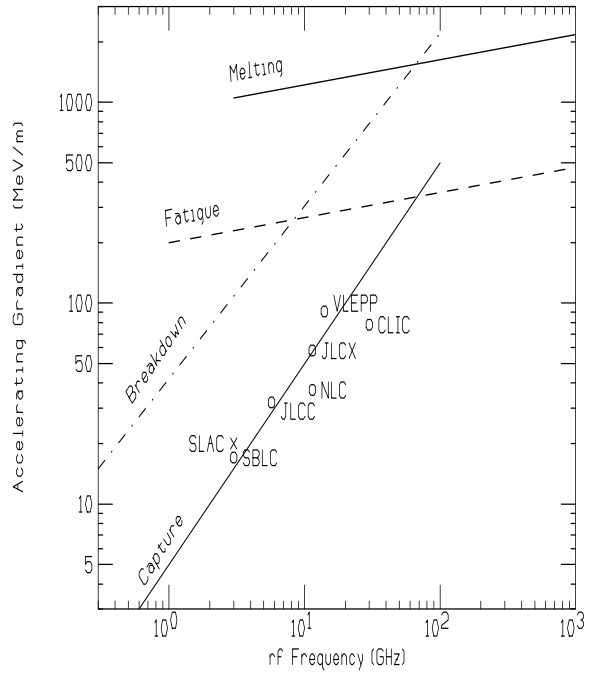,height=3.5in,width=3.0in}}
 \caption{Gradient values and limits in linear collider electron linacs} 
 \label{grad} 
 \end{figure} 
       If, however, the rf costs can be constrained, for instance when 
superconducting  cavities are used, then there will 
be no trade off and higher gradients should be expected to lower 
the length and cost. The gradients achievable in Niobium superconducting 
cavities is theoretically limited to about 40~MV/m and practically to 
15-25~MV/m. Nb$_3$Sn and high Tc materials may allow higher field gradients 
in the future with no loss of luminosity. 

The gradients for conventional structures 
have limits that are frequency dependent. Fig. \ref{grad} shows 
the gradient limits from breakdown, fatigue and dark current 
capture plotted against the operating rf frequency. Operating 
gradients and frequencies of several linear collider 
designs \cite{bluebook} are also indicated. One sees that the use 
of high frequencies allows higher accelerating gradients,
less overall length and thus, hopefully, less cost. There are 
however counterbalancing considerations from the requirements of 
luminosity. 


 The luminosity $\Ls$ of a linear collider can be written: 
 \b
   \label{lumeq}
   \Ls \= {1 \over 4\pi E}\ \ {N \over \sigma_x}\ \ {P_{\rm beam} \over \sigma_y}
   \ \ n_{\rm collisions}
 \e
where, in this case, $n_{\rm collisions}=1$; $\sigma_x$ and 
$\sigma_y$ are average beam spot sizes including any pinch 
effects: $\sigma_x$ being greater than $\sigma_y$; $E$ is the 
beam energy and $P_{\rm beam}$ is the total beam power. This can also 
be expressed as,
 \b
\label{constraint}
\Ls \= {1 \over 4\pi E}\ \ {n_\gamma \over 2 r_o \alpha}
       \ \ {P_{\rm beam} \over \sigma_y}
 \e
where $r_o$ is the classical electromagnetic radius, $\alpha$ is 
the electromagnetic constant, and $n_{\gamma}$ the 
number of photons emitted by one bunch as it passes through the 
other. If $n_\gamma$ is too large then the beamstrahlung 
background of electron pairs and other products becomes 
unacceptable. So, for a fixed criterion for background, we have: 
 $$
\Ls \propto {1 \over  E}\ \  {P_{beam} \over \sigma_y}
 $$
which may be compared to the required luminosity that increases 
as the square of energy, giving the requirement: 
 \b
\label{Ecubed}
{P_{\rm beam} \over \sigma_y}\ \propto \ E^3.
 \e
It is this requirement that makes it hard to design very high 
energy linear colliders. The 0.1 TeV SLC, with a relatively low 
energy, is still almost an order of magnitude below its design 
luminosity, and nearly four orders of magnitude less than that 
specified for the various designs for 
0.5 TeV linear colliders \cite{bluebook}.

\begin{figure}[ht!] 
\centerline{\epsfig{file=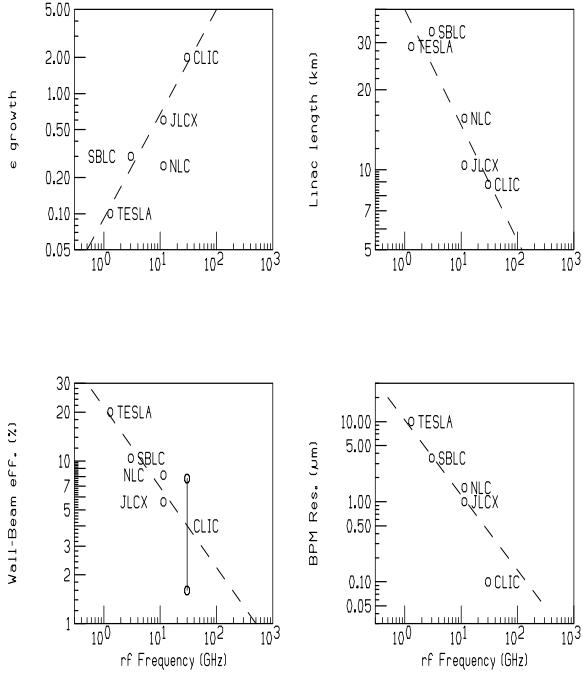,height=3.5in,width=3.0in}}
 \caption{Dependence of some sensitive parameters as a function of linear 
collider rf frequency.} 
 \label{freq} 
 \end{figure}

   Fig.\ref{freq}, using parameters from the linear collider 
proposals \cite{bluebook}, plots some relevant parameters against 
the rf frequency. One sees that as the frequencies rise, 
 \begin{itemize}
 \item the machine lengths fall as higher gradients become possible,
 \item greater alignment precision is required. For instance,
in the resolution of beam position monitors; and 
 \item despite these better alignments, the calculated  emittance growth 
during acceleration is greater; and
 \item the wall-power to beam-power efficiencies are less.
 \end{itemize}

Thus while length and cost considerations may favor high 
frequencies, yet luminosity considerations demand lower 
frequencies. 
 
\begin{figure}[ht!] 
\centerline{\epsfig{file=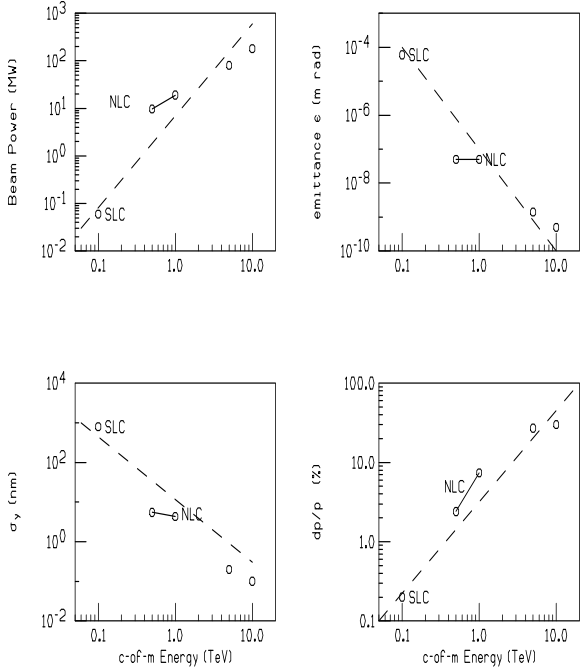,height=3.5in,width=3.0in}}
 \caption{Dependence of some sensitive parameters on linear collider energy. } 
 \label{energy} 
 \end{figure}

   At higher energies (as expected from Eq. \ref{Ecubed}),  
obtaining the required luminosity gets harder. Fig.\ref{energy} 
shows the dependency of some example machine 
parameters with energy.  SLC is taken as the example at 0.1 TeV, 
NLC parameters at 0.5 and 1 TeV, and 5 and 10 TeV examples are 
taken from a review paper by one of the authors \cite{me}. One sees 
that: 
 \begin{itemize}
 \item the assumed beam power rises approximately as $E^2$;
 \item the vertical spot sizes fall approximately as $E^{-2}$;
 \item the vertical normalized emittances fall even faster than $E^{-2}$; 
and
 \item the momentum spread due to beamstrahlung has been allowed to rise approximately 
linearly with $E$. 
 \end{itemize}

   These trends are independent of the acceleration method, 
frequency, etc, and indicate that as the energy and required 
luminosity rise, so the required beam powers, efficiencies, 
emittances and tolerances will all get harder to achieve. The 
use of higher frequencies or exotic technologies that would allow 
the gradient to rise, will, in general, make the achievement of 
the required luminosity even more difficult. It may well prove
impractical to construct linear electron-positron colliders, with 
adequate luminosity, at energies above a few TeV. 
                 
 \subsection{Photon-Photon Colliders}

   A gamma-gamma collider \cite{telnov} would use opposing electron 
linacs, as in a linear electron collider,  but just prior to the 
collision point, laser beams would be backscattered off the 
electrons to generate photon beams that would collide at the IP instead of the 
electrons. If suitable geometries are used, the mean 
photon-photon energy could be 80\% or more of that of the 
electrons, with a luminosity about 1/10th. 

   If the electron beams, after they have Compton backscattered the 
photons, are deflected, then backgrounds from beamstrahlung can 
be eliminated. The constraint on ${N / \sigma_x}$ in Eq.\ref{lumeq} is thus removed and one might hope that higher 
luminosities would now be possible by raising $N$ and lowering 
$\sigma_x$. Unfortunately, to do this, one needs sources of 
larger number of electron bunches with smaller emittances, and one must 
find ways to accelerate and focus such  beams without excessive 
emittance growth. Conventional damping rings will have difficulty 
doing this \cite{mygamma}. Exotic electron sources might be needed. 

   Thus, although gamma-gamma collisions can and should be made 
available at any future electron-positron linear collider, to add 
physics capability, they may not give higher luminosity for a 
given beam power. 
\subsection{Muon-Muon Colliders}
   There are two advantages of muons, as opposed to 
electrons, for a lepton collider. 
 \begin{itemize}
\item
   The synchrotron radiation, that forces high energy electron 
colliders to be linear, is (see Eq. \ref{syncheq}) inversely 
proportional to the fourth power of mass: It is negligible in muon 
colliders with energy less than 10 TeV. Thus a muon collider, up 
to such energy, can be circular. In practice this means in 
can be smaller. The linacs for a 0.5 TeV NLC would be 20 km 
long. The ring for a muon collider of the same energy would be 
only about 1.2 km circumference. 

\item
   The luminosity of a muon collider is given by the same formula as in Eq. \ref{lumeq} as given above for an electron positron collider, but there are 
two significant changes: 1) The classical radius $r_o$ is now that for the muon 
and is 200 times smaller; and 2) the number of collisions a bunch can make
$n_{collisions}$ is no longer 1, but is now related to the average bending 
field in the muon collider ring, with
 $$
n_{collisions} \ \approx \ 150  \ B_{ave}
 $$
With an average field of 6 Tesla, $n_{collisions}\approx 900$.
Thus these two effects give muons an {\it in principle} luminosity 
advantage of more than $10^5$. 
 \end{itemize} 

   The problems with the use of muons are:
 \begin{itemize}
 \item Muons can be best obtained from 
the decay of pions, made by higher energy protons impinging on a target.
A high intensity proton source is thus required and  
very efficient 
capture and decay of these pions is essential.
 \item Because  the muons are made with very large emittance, 
they must be cooled and this must be done very rapidly because of their short lifetime. Conventional synchrotron, electron, or stochastic cooling is
too slow. Ionization cooling is the only clear 
possibility, but does not cool to very low emittances.
 \item Because of their short lifetime, conventional 
synchrotron acceleration would be too slow. Recirculating 
accelerators or pulsed synchrotrons must be used.
 \item
Because they decay while stored in the collider, muons radiate
the ring and 
detector with their decay products. Shielding is essential and backgrounds
will certainly be significant.
 \end{itemize}

 \begin{figure}[t!] 
\centerline{\epsfig{file=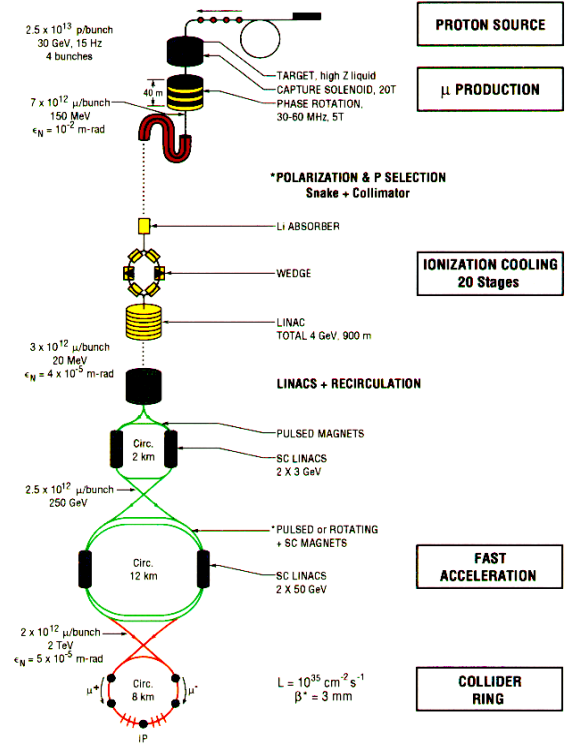,height=4.0in,width=3.0in}}
 \caption{Overview of a 4 TeV Muon Collider} 
 \label{overview}
 \end{figure} 
Muon colliders were first considered more than 20 years ago,  many papers have
been written and many workshops  held \cite{history}. A collaboration, lead by
BNL, FNAL and LBNL,  with contributions from 18 institutions has been studying a
4 TeV, high luminosity  scenario and presented a Feasibility Study \cite{book}
to the  1996 Snowmass Workshop. 

The basic parameters of this collider are shown 
schematically in Fig.\ref{overview} and given in 
Tb.\ref{sum} together with those for a 0.5 TeV 
demonstration machine based on the AGS as an injector. It 
is assumed that a demonstration version based on upgrades 
of the FERMILAB, or CERN machines would also be possible. 

\begin{table}[thb!]  
\caption{Parameters of Collider Rings}
\label{sum}
\begin{tabular}{|l|lc|c|}
\hline
c-of-m Energy              & TeV      & 4     & .5       
\\
\hline
Beam energy                & TeV      &     2    &   .25 
   \\ \hline
Beam $\gamma$              & $10^3$    &   19 &  
2.4   \\ \hline
Repetition rate            & Hz       &    15    &    
2.5    \\ \hline
Muons per bunch            & $10^{12}$  &   2    &    4  
    \\  \hline
Bunches of each sign       &          &   2      &    1  
    \\   \hline
Norm. {\it rms} emit.   $\epsilon_N$   &$\pi$\, m\,rad & 
 $510^{-5}$  &  $910^{-5}$    \\ \hline
Bending Field              &  T   &    9    &    9     
\\  \hline
Circumference              &  Km      &    7    &    1.2 
   \\  \hline
Ave. ring  field $B$    & T   & 6    &   5       \\ \hline
Effective turns          & $10^2$   & 9    &   8   \\ \hline
$\beta^*$ at intersection   & mm     &   3   &   8     
\\   \hline
{\it rms} I.P. beam size        & $\mu m$&   2.8 &  17   
 \\  \hline
Luminosity &${\rm cm}^{-2}{\rm s}^{-1}$& 
$10^{35}$&$10^{33}$\\  \hline
\end{tabular}
\end{table}
The main components are: 
\begin{itemize}
\item  A proton source with KAON like parameters (30 GeV, $10^{14}$ 
protons per pulse, at 15 Hz). 

\item A liquid metal target surrounded by a 20~T hybrid or high T$_c$ 
superconducting  solenoid to make and capture pions. 

\item A 5 T solenoidal channel within a sequence of rf cavities is 
used to allow the pions to decay into muons and, at the same 
time, decelerate the fast ones that come first, while 
accelerating the lower momentum ones that come later. Muons from 
pions in the 100-500 MeV range emerge in a 6 m long bunch at 150 
$\pm$ 30 MeV bunch. 

\item A solenoidal snake and collimator to select 
the momentum, and thus polarization, of the muons. 

\item A sequence of 20 ionization cooling stages, each consisting of 
a) lithium energy loss rod in a strong focusing environment for 
transverse cooling, b) linac reacceleration and c) lithium wedges in a 
dispersive environment for cooling in momentum space. 

\item A linac, and/or recirculating linac, pre accelerator, followed by
a sequence of pulsed field synchrotron accelerators using 
superconducting linacs for rf.

\item An isochronous collider ring with locally corrected low beta 
($\beta$=3 mm) insertion.
\end{itemize}

   For a low energy muon collider, there would be a relatively large fixed cost for the 
muon source, but for a high energy machine the cost would still be dominated 
by that of the final circular accelerator and collider rings. Estimates suggest that 
the cost of these might be as much as a factor 3 higher than that for a hadron 
machine of the same beam energy, but, because of the advantage in colliding 
point like leptons, a factor of 3 or more less than a hadron machine of the 
same {\it effective energy}. 
\subsection{Comparison of Machines}
 \begin{figure}[hbt!] 
\centerline{\epsfig{file=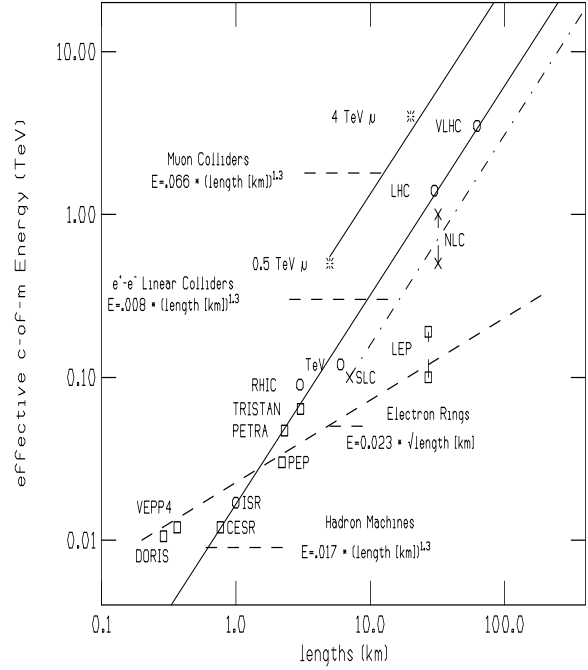,height=3.5in,width=3.0in}}
 \caption{Effective energies of colliders as a function of their total 
length.} 
 \label{length} \end{figure}

   In Fig. \ref{length}, the effective energies (as defined by 
Eq. \ref{Eeffeq}) of representative machines 
are plotted against their total tunnel lengths. We note:

 \begin{itemize}
 \item  Hadrons Colliders:
 It is seen that the energies of machines rise with their size, but 
that this rise is faster than linear ($E_{\rm eff}\propto L^{1.3}$). This
slope is a reflection in the steady rise in bending magnetic 
fields used as technologies and materials have become available. 

 \item Circular Electron-Positron Colliders:
The energies of these machines rise approximately as the square root of their 
size, as expected from the cost optimization discussed above.

 \item Linear Electron-Positron Colliders:
The SLC is the only existing machine of this type and only one example of a 
proposed machine (the NLC) is plotted. The line drawn has the same slope as 
for the hadron machines and implies a similar rise in accelerating 
gradient, as technologies advance. 

 \item Muon-Muon Colliders:
Only the 4 TeV collider, discussed above, and the 0.5 TeV {\it  demonstration
machine}  have been plotted. The line  drawn has the same slope as for the
hadron machines. 
\end{itemize}

 \begin{figure*}[t!] 
\centerline{\epsfig{file=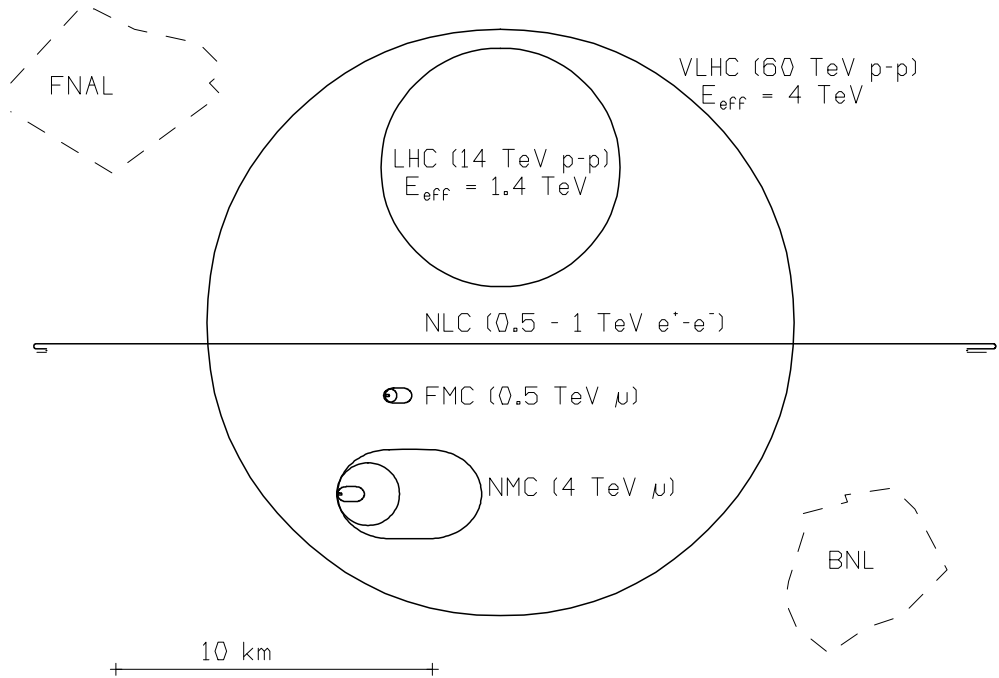,height=3.52in,width=5.18in}}
 \caption{Approximate sizes of some possible future colliders.} 
 \label{examples} \end{figure*} 

 \begin{figure}[t!] 
\centerline{\epsfig{file=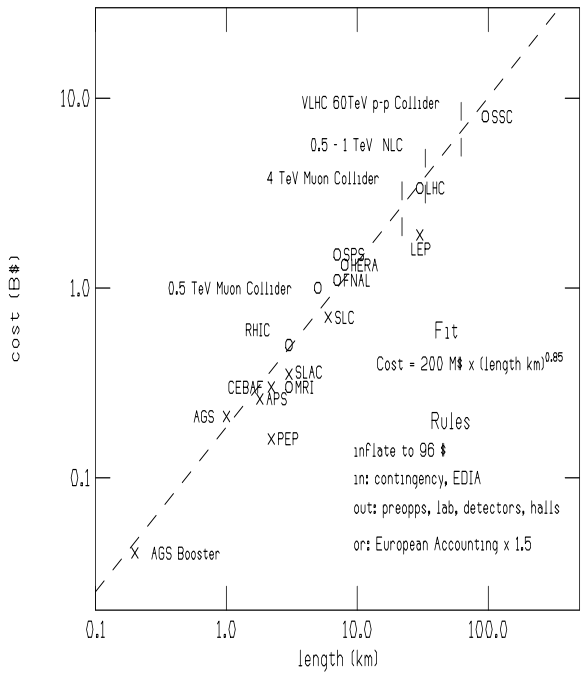,height=3.5in,width=3.0in}}
 \caption{Costs of some machines as a function of their total lengths.}
 \label{cost} \end{figure}

   It is noted that the muon collider offers the greatest energy per unit 
length. This is also apparent in Fig. \ref{examples}, in which the 
footprints of a number of proposed machines are given on the same scale. But 
does this mean it will give the greatest energy per unit of cost ? 
 Fig. \ref{cost} plots the cost of a sample of machines 
against their size. Before examining this plot, be warned:
the numbers you will see will not be the ones you are familiar 
with. The published numbers for different projects use different 
accounting procedures and include different items in their costs. 
Not very exact corrections and escalation have been made to obtain 
estimates of the costs under fixed criteria: 1996 \$'s, US 
accounting, no detectors or halls. The resulting numbers, as 
plotted, must be considered to have errors of at least $\pm$ 20\%. 

The costs are seen to be surprisingly well represented by a 
straight line. Circular electron machines, as expected, lie 
significantly below this line. The only plotted muon collider 
(the 0.5 TeV demonstration machine's very preliminary cost 
estimate)  lies above the line. But the clear indication 
is that length is, or at least has been, a good estimator of 
approximate cost. It is interesting to note that the fitted line 
indicates costs rising, not linearly, but as the 0.85\,th power of 
length. This can be taken as a measure of economies of scale. 
\section{Conclusion}
Our conclusions, with the caveat that they are indeed only our opinions, 
are: 
 \begin{itemize}
 \item   The LHC is a well optimized and appropriate next step towards high 
{\it effective} energy.

\item   A Very Large Hadron Collider with energy greater than the 
SSC (e.g. 60 TeV c-of-m) and cost somewhat less than the SSC, may 
well be possible with the use of high T$_c$ superconductors that 
may soon be available. 

\item   A ``Next Linear Collider" is the only clean way to 
complement the LHC with a lepton machine, and the only way to do 
so soon. But it appears that even a 0.5 TeV collider will be more 
expensive than the LHC, and it will be technically 
challenging: obtaining the design luminosity may not be easy. 

\item Extrapolating conventional rf \ee linear colliders to 
energies above 1 or 2 TeV will be very difficult. Raising the rf 
frequency can reduce length and probably cost for a given energy, 
but obtaining  luminosity increasing as the square of energy, as 
required, may not be feasible. 

\item   Laser driven accelerators are becoming more realistic and 
can be expected to have a significantly lower cost per TeV. But 
the ratio of luminosity to wall power and the ability to preserve very small 
emittances, is likely to be significantly worse than for 
conventional rf driven machines. Colliders using such technologies are thus 
 unlikely to achieve very high luminosities and thus unsuitable for 
higher (above 2 TeV) energy physics research. 

\item A higher gradient superconducting Linac collider using Nb$_3$Sn or
high T$_c$ materials, if it becomes technically possible, could be the only 
way to attain the required luminosities in a higher energy \ee collider.

\item   Gamma-gamma collisions can and should be obtained at any 
future electron-positron linear collider. They would add physics 
capability to such a machine, but, despite their freedom from the 
beamstrahlung constraint, are unlikely to achieve higher 
luminosity.

\item   A muon Collider, being circular, could be far smaller 
than a conventional electron-positron collider of the same 
energy. Very preliminary estimates suggest that it would also be 
significantly cheaper. The ratio of luminosity to wall power for 
such machines, above 2 TeV, appears to be better than that for 
electron positron machines, and extrapolation to a center of mass 
energy of 4 TeV or above does not seem unreasonable. If research 
and development can show that it is practical, then a 
0.5-1 TeV muon collider could be a useful complement to \ee colliders, and, 
at higher energies (e.g. 4 TeV), could be a viable alternative. 
   \end{itemize} 


\section*{References}
\begin{thebibliography}{9}
 \bibitem{lumlim}A. W. Chao, R. B. Palmer, L. Evans, J, Gareyte, R. H. 
Siemann, {\it Hadron Colliders (SSC/LHC)}, Proc.1990 Summer Study on High 
Energy Physics, Snowmass, (1990) p 667.
 \bibitem{magcost} R. B. Palmer, {\it to be published}.
 \bibitem{pipe} S. Holmes for the RLHC Group, {Summary Report}, presentation at the Snowmass Workshop 1996, to be published.
 \bibitem{bluebook}{\it International Linear Collider Technical Review 
Committee Report}, SLAC-R-95-471, (1995)
 \bibitem{me}R. B. Palmer,{\it Prospects for High Energy \ee Linear 
Colliders}, Annu. Rev. Nucl. Part. Sci. (1990) 40, p 529-92.
 \bibitem{telnov} V. Telnov, Nucl. Instr. and Meth. A294, (1990) 72; {\it A Second Interaction Region for Gamma-Gamma, Gamma-Electron and Electron-Electron Collisions for NLC}, Ed. K-J Kim, LBNL-38985, LLNL-UCRL-ID 124182, SLAC-PUB-95-7192.
 \bibitem{book}{\it $\mu^+\mu^-$ collider, A Feasibility Study}, BNL-52503,FermiLab-Conf-96/092, LBNL-38946, submitted to the Proceedings of the Snowmass96 Workshop.
 \bibitem{mygamma} R. B. Palmer,{\it Accelerator parameters for $\gamma-
\gamma$ colliders}; Nucl. Inst. and Meth., A355 (1995) 150-153.

 \bibitem{history}A. N. Skrinsky and V.V. Parkhomchuk, Sov. J. of
Nucl. Physics {\bf 12}, (1981) 3; 
  Neuffer, IEEE Trans. {\bf NS-28}, (1981) 2034;
  E. A. Perevedentsev and A. N. Skrinsky, Proc. 12th Int. Conf. on High Energy 
Accelerators, F. T. Cole and R. Donaldson, Eds., (1983) 485; 
  {2$^{nd}$ Workshop}, Sausalito, CA, Ed.
D. Cline, AIP Press, Woodbury, New York, (1995).
\end{thebibliography}
\end{document}